\documentclass[sigconf]{acmart}
\usepackage{booktabs}
\usepackage{multirow}

\def\BibTeX{{\rm B\kern-.05em{\sc i\kern-.025em b}\kern-.08emT\kern-.1667em\lower.7ex\hbox{E}\kern-.125emX}}
    
\copyrightyear{2019}
\acmYear{2019}
\setcopyright{acmcopyright}
\acmConference[SMSociety '19]{International Conference on Social Media and Society}{July 19--21, 2019}{Toronto, ON, Canada}
\acmBooktitle{International Conference on Social Media and Society (SMSociety '19), July 19--21, 2019, Toronto, ON, Canada}
\acmPrice{15.00}
\acmDOI{10.1145/3328529.3328541}
\acmISBN{978-1-4503-6651-9/19/07}

\begin{document}

%

 \title{A Socio-Informatic Approach to Automated Account Classification on Social Media}

%

\author{Laurenz A. Cornelissen}
\affiliation{%
  \institution{Computational Social Science Group \\ Centre for AI Research \\ Department of Information Science \\ Stellenbosch University}
  }
\email{alducornelissen@sun.ac.za}

\author{Petrus Schoonwinkel}
\affiliation{%
  \institution{Computational Social Science Group \\ Centre for AI Research \\ Department of Information Science \\ Stellenbosch University}}
\email{petrus.schoonwinkel@gmail.com}

\author{Richard J Barnett}
\affiliation{%
  \institution{Computational Social Science Group \\ Centre for AI Research \\ Department of Information Science \\ Stellenbosch University}}
\email{barnettrj@acm.org}

%

 \renewcommand{\shortauthors}{Laurenz A. Cornelissen, et al.}

%
\begin{abstract}
Automated accounts on social media have become increasingly problematic. We propose a key feature in combination with existing methods to improve machine learning algorithms for bot detection. We successfully improve classification performance through including the proposed feature.
\end{abstract}

%
%

%

%

%
\maketitle

\section{Introduction}

Automated accounts on social networking sites (SNS) are increasingly problematic, due to their use in manipulating political and social issues online. Many observers would recognise that this is simply a form of marketing, i.e. intentionally influencing opinions on topics, products and services. However, the use of these tactics are perceived to be more unethical when applied to political and social environments.

As early as 2011 \citep{Lee2011}, researchers started developing techniques to identify malicious or dishonest accounts on SNS. Early approaches were effective at identifying simple automated accounts. The success of subsequent methods are predominantly due to more data and the continuous improvement in machine learning capabilities. In time, however, automated accounts became more sophisticated, necessitating more advanced detection methods.

The majority of features included in machine learning models rely on profile, temporal, content, and behavioural aspects of accounts. This study proposes a novel feature, which we call a socio-informatic feature, to extend existing classification models. We propose that the inclusion of the socio-informatic feature into existing classification methods would improve the classification performance of existing approaches.

\subsection{Political Interference}

The recent increase in political interference by automated accounts is an exogenous motivation for automation classification. The problem of automated SNS accounts is not new, but, it has mostly been an annoyance to the operators of the platform. Recently, there have been potentially serious consequences of such interference on these platforms, driving the motivation to improve classification methods. Improved classification methods would root out automated accounts improve trust in honest users in this social media age. 

\citet{Hemsley2018} highlights several noteworthy political events since 2016 that were suspected of being influenced by automated accounts. As an example, a fifth of the users participating in the online Twitter discussion were suspected to be automated agents during the 2016 United States presidential election \citep{Ferrara2016}. 

These prominent political interference campaigns accelerated the need to investigate the political interference on SNS. There is mixed evidence of the level to which political interference on SNS have an actual impact. Regardless, there is clear evidence of attempts at political interference \citep{Rossini2018, Ross2019}.

\subsection{Automated Account Classification}

An early attempt at classifying automated accounts was done by \citet{Lee2011}. They deployed 60 honeypots on Twitter for seven months, which attracted interactions from 36000 candidate accounts. However, the honeypot approach to bot classification,\footnote{We use automated agent and bot interchangeably.} is inherently limited in the manner by which the candidate accounts are attracted  \citep[see][]{Cornelissen2018}.

Another early approach involved using human annotators to classify accounts. Comparing experts and ``turkers'', \footnote{``Turkers'' is a colloquial term used to describe rented labour for Amazon Mechanical Turk.} \citet{Wang2012} found that experts are near-optimal in classification effectiveness. This method is costly, especially considering that automated accounts can easily be created quickly in the thousands.

Subsequent studies on bot classification are dominated by computational methods, more specifically, machine learning. As an example of unsupervised machine learning, \citet{Miller2014}, used clustering algorithms to detect automated accounts from various account features. As an example of a supervised approach, \citet{Ji2016} used random forest classification on 18 identified features. More creative approaches, such as those by \citet{Cresci2016}, discriminates between genuine and automated accounts by using a DNA analysis inspired technique.

In 2015 DARPA organised a competition in response to an increase in the problem of automated accounts on social media. Six teams competed in the challenge and the better performing classification methods relied on Tweet syntax and semantics, temporal features, as well as user profile and interaction data \citep{Subrahmanian2016}. These features are becoming all but standard practice in the automated account detection research domain. Most subsequent approaches use varying subsets of the same features on which to train machine learning libraries.

This study intends to join the research in developing better detection techniques, but opts to focus on the value of the underlying data features in the pursuit of improving bot classification. Instead of increasing the number of features in the training data, and testing increasingly advanced machine learning methods, we propose that an deliberated and contextualised understanding of what is being classified would help in producing more productive features.

\section{Background Literature}
Just as the need to sleep lead researchers to identify circadian rhythms as indicative of non-human behaviour, socially driven actions may offer a robust signal of human activity. The objective is, then, to find a measure of socially driven actions, which are difficult or costly to imitate by automated or poor social agents. Many actions on social media are inherently social, such as liking or sharing content, so observing the patterns in these actions should offer insights as to the sociability of an account. A key social action, follower networks, is missing as a feature in current literature, which may offer an improvement in the detection of true social agents. 

To explain the novelty and benefit of the proposed features, there are two factors to consider. The first is from information load literature, particularly from information channel repertoire literature \citep{Liang2017, Kim2016, Wolfsfeld2016}; the second utilises the investigation of social network structure, which is well explored in the field of social network analysis \citep{Wasserman1994}. In simple terms, users are considerate of who they follow due to the risk of information overload, which results in particular network patterns. Social network analysis, then, offers the tools to be able to analyse these resulting patterns.

\subsection{Information Repertoire}
A key contributor to following behaviour online is bounded human attention, particularly avoiding the risk of information overload \citep{Liang2017}. When an individual follows another on a SNS, they subscribe to observe the followed account's actions and broadcasts (tweets), which increases exposure to information. Over-subscription to other accounts would, therefore, contribute to information overload. In contrast, actions such as replying, retweeting and liking do not contribute to the information load of the user doing the action. According to \citep{Liang2017} and \citep{Kim2016} people therefore tend to rely on a small or manageable repertoire of information channels. Over time, users tend to curate a consistent number of channels, removing redundant channels to optimise information load. In other words, actions like tweeting, retweeting and replying (producing content) does not contribute to the user's information load, it does contribute to their followers' information load. Consuming information, by following others, does however contribute directly to the user's information load. Following behaviour could, therefore, be seen as a more deliberated social action.

The above would imply that certain following behaviour, or strategies, would signal certain characteristics of the user. As an example in a political context, individuals with \emph{richer} curated information repertoires are found to have higher levels of political knowledge, efficacy, and participation compared to those with less \emph{rich} information sources \citep{Wolfsfeld2016}. The implication is that more considered curation of follower networks (information repertoire) is indicative of social agents who deliberate more about their following behaviour. It is na\"{i}ve to assume that active curating invariably leads to quality information repertoires, since \emph{quality} is ill-defined. There is ample evidence of users curating self-defeating echo-chambers \citep{Adamic2005}. Nevertheless, even a repertoire resembling an echo-chamber has social motivations, such as homophily \citep{Kossinets2009}.

Regardless of the information curating strategy, and its effectiveness for information load optimisation, the following patterns of online users reveal many other social preferences. The information load narrative considers a monadic social entity, whereas the next section considers these actions on a dyadic and triadic level.

\subsection{Social Network Theory}
The previous section described a user's following actions as a function of information load considerations for the individual. To uncover the social considerations of users, we need to reveal the social structure of their actions. A simple approach is to investigate the actions on a dyadic and triadic level. For example: how many of the followed users follow back? (dyadic); and, how many of the followed users follow each other? (triadic). Assuming that information load is not an issue for the user deciding to follow another, there are still social considerations for following actions. The decision to follow a person back, when they do not necessarily contribute valuable information, is an hierarchical social choice. 
For example, if A follows B, and B does not follow back, yet A continues following B, B has a higher hierarchical social position than A. If B did follow back, they have equal standing. If B does not follow back, and A un-follows B after a while, they also have equal social status \citep[see][]{Everett2012}. On the triadic level, the dynamics become more complicated. On a SNS platform such as Twitter, where following is one directional, there are 16 possible combinations of triads, compared to three in a dyad \citep[][p.556]{Wasserman1994}. Consider again the social dynamics between three actors, A, B and C. If A follows B, and B follows C the transitive principle implies that A should also follow C. The need of individuals in A's position to \emph{close} the triad is a well documented psychological need \citep[see][]{Flynn2010}, it is well observed as a key contribution to large scale social network dynamics \citep{Kossinets2009}, and such triadic based considerations of individuals are key to prominent social network theory \citep{Burt2005,Granovetter1973}. Individual behaviour is, therefore, influenced by structural constraints of their network and the relation between units become the focus of analysis, instead of the unit itself.

\subsection{Socio-informatic Feature}
We call this approach \emph{socio-informatic} since it captures two aspects of following behaviour for humans: social and information. Although people can get information from non-human sources such as books, newspapers, or Wikipedia, they still rely on particular individuals or groups of individuals as key information sources. This is the particular appeal of a SNS such as Twitter, where one can directly follow an individual, such as a celebrity or journalist, instead of relying on relayed information. It is also difficult, if not impossible, to separate the social aspects of such information seeking activities. There are many social considerations people employ in their curation of their information sources, which is key to many observed phenomena such as echo-chambers, where people curate information due to more social reasons than informational. We do not propose that we capture the full rich aspects of a socio-informatic approach to user profiling, but we offer this as a first step towards investigating how this might aid in improving the identification of honest and good quality social agents among a flood of automated and semi-automated accounts.

A central assumption of this approach is that the consideration for following others on SNS is deliberate enough to be able to distinguish a well functioning social agent (honest human) from that of a automated account. Analysing these more expensive social actions produces a set of variables that can be used to create a model to classify the features of a given user as an automated account or an honest human. Almost all previous attempts have focused on what can be described as less socially expensive actions \citep[see][]{Subrahmanian2016}. The next two sections expand on each motivation.

A user could reasonably follow and link to any source which they judge as a worthy source of information, but unfettered linkage would lead to information overload for the user \citep{Liang2017}. Thus, part of the following motivation is for informational reasons. However, social considerations also play a key role in the motivation for following others. Individuals follow others that are similar to themselves due to homophily, and this is also observed on SNS \citep{Aiello2012}. Moreover, there are generic social rules which dictates link formation on any social network, particularly preferential attachment \citep{Barabasi1999}. The action of following others on SNS should therefore resemble a non-random pattern of behaviour. 

One way to operationalize these patterns is to construct a network representation from it. The exact network measures will be discussed in Section \ref{Sec:NetworkMeasures}. As an example, we discuss two measures here: centrality and clustering coefficient. Centrality represents the degree of importance of a node in a graph. There are many measures of centrality, but the simplest measure is to count the number of connections to a node. Given that a node has a higher amount of relations than any other node in the graph, it will have the highest degree centrality. Clustering coefficient is a indicative measure of how transitive a relation is between any three nodes. In other words, it is a way to measure the social dynamic mentioned above, whether \textit{a friend of a friend, is} also \textit{a friend}. 

\section{Research Method}

The objective of this paper is to improve the prediction of SNS accounts as automated or not, by including a novel social feature of the accounts. Key to the novel feature is the fact that it should capture an individual's network repertoire as well as social structure. To do this, we have three needs. First, to test the accuracy of any prediction, we need a list of Twitter users which have been annotated with a ground truth of bot or not. Second, for each of these accounts, we need to capture their information repertoire. Lastly, to predict whether an account is automated, based on the information repertoire feature, a machine learning approach could be used, since it has been the superior approach in past research. In this section we will elaborate on the three needs and specify at each step why that method is used. 

\subsection{Data Collection}

To obtain a list of annotated Twitter users, we used the publicly available data provided by \citet{Varol2017}. The data consists of a Twitter user ID and a manually annotated binary label, indicating whether it is a bot or not. The accounts in the varol dataset were re-crawled from Twitter in March 2018, and again in January 2019, to determine if the dataset still contained a majority percentage of the original users. This confirmed that of the original 2573 accounts, 2178 were still active in March 2018, and 2133 were still accessible in January 2019. Of the accessible accounts, 33\% (671) labelled as a bot and 67\% (1462) as not a bot. In order to measure the performance of our classifier against a third party classifier, we use the scores of each account, provided by Botometer.\footnote{Accessible at \url{https://botometer.iuni.iu.edu}.} Botometer is a widely accepted industry standard for bot classification and provided the baseline for this project. While it is possible to reproduce other approaches to bot classification, most of these methods are either contained as an ensemble approach in Botometer, or outside the scope of this project. In Section \ref{Sec:slmodels}, we will discuss how the various scores are used to train different models. Given the identifiers, it is possible to crawl their respective follower networks. 

\begin{figure}
    \centering
    \includegraphics[width=\linewidth]{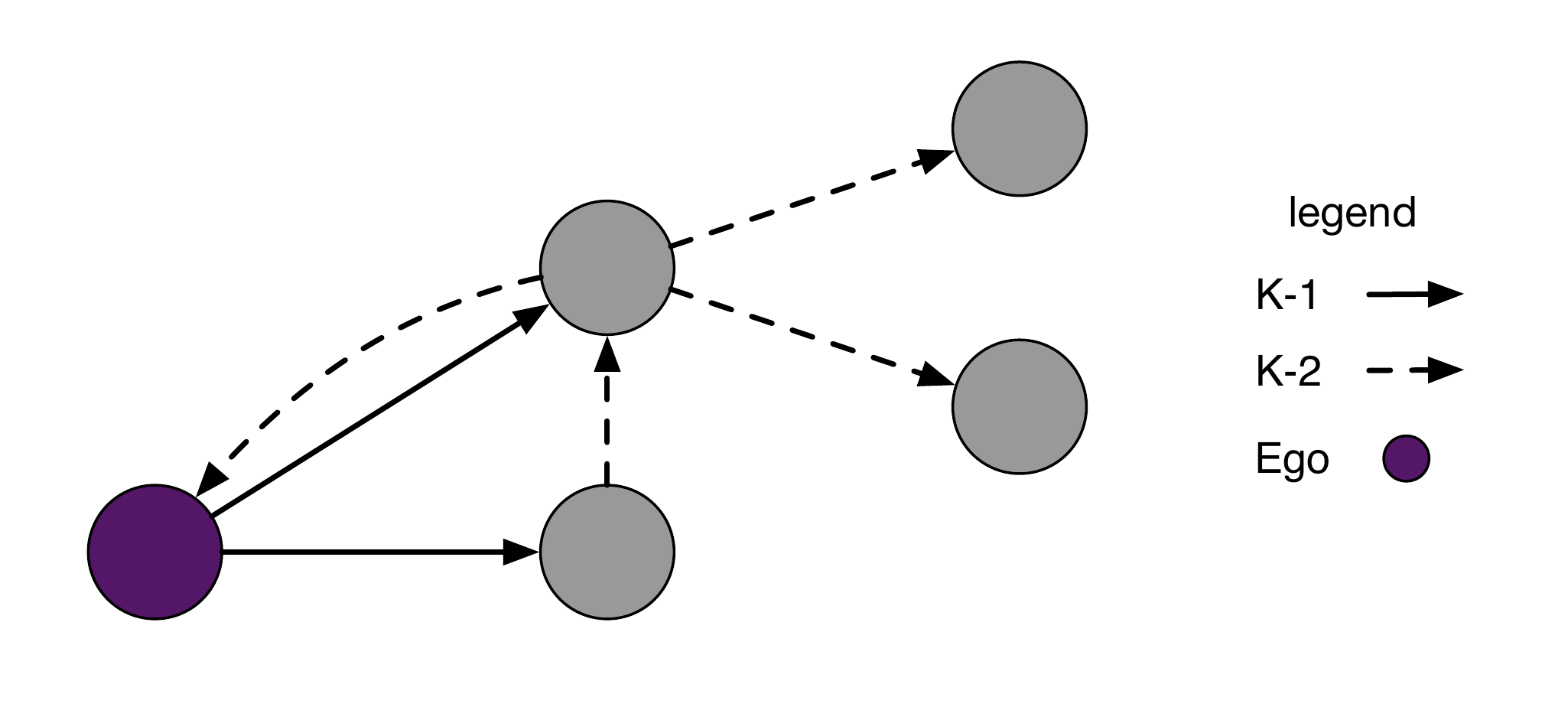}
    \caption{Descriptive Diagram of k-2 Crawl}
    \label{fig:k2crawl}
\end{figure}

For each account in the list we need to capture the information repertoire, which would involve capturing who they follow. However, capturing just a list of followers is not enough to construct a social network on a triadic level. We therefore need to also capture from this list, those they follow in turn. We queried the Twitter API and obtained the follower network, two steps from each identified Twitter user. A third step would be beneficial, but impractical, since this could theoretically gather the whole Twitter network with a median distance of 4.12 reached in 2010 \citep[][594]{Kwak2010}. Figure~\ref{fig:k2crawl} illustrates this network structure. 

Utilising at the follower network as a feature is not a new approach. For instance, \citet{Yang2013} gathered the friends and followers of each user. However, in terms of capturing information repertoire of a user, gathering their friends is unnecessary. Moreover, a single step of users the account follows offers no information of relations on a triadic level.

The network is then reduced by means of a k-core decomposition to produce an additional view of the core network of the ego. Such a reduced ego-centric graph may be more reflective of the user's socio-informatic features by reducing unnecessary noise \citep{Alvarez2005}. For the purposes of this analysis, both the full and the reduced networks will be used in the ensemble, since each captures unique features. 

With necessary data gathered, it is required to prepare the data, since the social network data is not amenable in it's raw form for most machine learning approaches. 

\subsection{Data Preparation}

The data is not suitable for analysis in its raw form, which necessitates pre-processing. There are many ways to prepare data for machine learning algorithms. For the purpose of this analysis, a set of network measures will be performed on the data in order to reduce the varied size of the network and make it readable for the machine learning library. 

\subsubsection{Social Network Measures}
\label{Sec:NetworkMeasures}

Most supervised machine learning algorithms require input tensors to be of uniform size, whereas individual networks provide inconsistently sized two-dimensional tensors. As a result, a series of measurements must be performed on the networks, in order to transform them into consistent input. The measurements capture the complexity of the information repertoire of each user, and will be used to create the classification models. 

The following network measures were calculated for each observation in the dataset: number of vertices, number of edges, global clustering coefficient, local clustering coefficient, centralisation in-degree, centralisation out-degree, centralisation degree, centrality in-degree, centrality out-degree and centrality degree, network density, reciprocity, assortativity and articulation points. In addition to these measures, the three vertex centrality measures were also normalised and included in the set. The various measures are described in Table~\ref{T:Network_Measures}, and are largely based on those used in \citep{Cornelissen2018b}. These measures can now be used as part of an ensemble learning model to classify users. 

\subsubsection{Data Split}
A common first step for many supervised machine learning methods is to split the dataset into a training and test set. It is also beneficial to scale the data before passing it to machine learning algorithms. We split the data through an 80/20 split of training and testing sets using a random index. The training set consists of 1367 observations and the test set has 342 observations. The training set contained 398 (29.2\%) observations labelled as bots, and 969 observations (70.8\%) labelled as not. Similarly, the test set has 112 (32.8\%) labelled as bots, and 230 (67.2\%) labelled as not. This indicates that both the training and test sets are approximate representative samples of the original full dataset. After the data was split into training and test sets, the predictor sets were scaled using the FastScale function from the dataPreparation package by \citet{Toulemonde2019}.

\begin{table*}[]
\centering
\footnotesize
\caption{Network Measures}
\label{T:Network_Measures}
\begin{tabular}{@{}p{0.2\linewidth}p{0.65\linewidth}p{0.15\linewidth}@{}}
\toprule
\textbf{Measure}				& \textbf{Description} 							& \textbf{Source}\\ \midrule
Number of Vertices 							& The number of vertices in the graph. & \citep[p.5-8]{Harris2008}\\
Number of Edges 							& The number of edges in the graph. & \citep[p.5-8]{Harris2008}\\
Global Clustering Coefficient  & ``The fraction of paths of length two in the network that are closed''. i.e. Whether a friend of a friend is a friend. & \citep[p.~199]{Newman2010}    \\
Local Clustering Coefficient  & The same as the global measure, but measured for a focal node. & \citep[p.~201]{Newman2010} \\
In-degree Graph Centralisation & A graph-level measure of the number of edges directed towards the nodes in a graph.& \citep[p.~175-177]{Wasserman1994}\\
Out-degree Graph Centralisation & A graph-level measure of the number of edges directed from the nodes in a graph.& \citep[p.~175-177]{Wasserman1994} \\
Degree Graph Centralisation   & A graph-level measure of the number of edges directed to and from the nodes in a graph.& \citep[p.~175-177]{Wasserman1994}\\
In-degree Graph Centrality   & The number of edges that are directed towards a single node.& \citep[p.~178,199-202]{Wasserman1994}\\
Out-degree Graph Centrality   & The number of edges that are directed from a single node to other nodes in the graph.& \citep[p.~178,199-202]{Wasserman1994}\\
Degree Graph Centrality     & The number of edges directed from and at a single node in a graph.& \citep[p.~178,199-202]{Wasserman1994}\\
Density             & Ratio of the amount of edges and the amount of possible edges in the graph.& \citep[p.~165]{Wasserman1994}\\
Reciprocation          & The proportion of mutual connections in a directed graph.& \citep[p.~515]{Wasserman1994}\\
Assortativity          & Also known as assortative mixing. Assortativity is the preference for a graph's nodes to attach itself to other nodes that are similar to it. The similarity, in this case is measured by degree. & \citep{Newman2003}\\
Articulation Points       & Nodes that if removed, increase the number of connected components in a graph. Also known as cut vertices.& \citep{Italiano2012}                                                                      \\ \bottomrule
\end{tabular}
\\\vspace*{0.5\baselineskip}\footnotesize \emph{Source: \citet{Cornelissen2018b}.}
\end{table*}

\subsection{Ensemble Learning}
\label{Sec:slmodels}
For this paper we used R version 3.5.2, to analyse the data. The choice of ensemble learning approach is directed by ease of use and performance with broad application areas. We chose to use the Super Learner (SL) algorithm for ensemble learning, which is a stacking ensemble approach, formalised  by \citet{VanderLaan2007}. Stacking ensemble learning is simply the act of stacking multiple base learners (each of which could be an ensemble itself) in order to produce better results than any one of the base learners. The SL algorithm thus requires a specification of a library of appropriate base learners. Each base learner is trained on the same training set, and the V-fold cross validated results are gathered into a matrix which is used by the SL algorithm as a training set to create an ensemble model from the predictions. We opted for the SuperLearner package by \citet{Polley2018}, which offers a decent standard library of learners and is well documented.

\subsubsection{Variable Specification}

The ensemble learning algorithm requires a dataset of observations, which consist of input vectors, i.e. network measures; and outcome, i.e. labels. Table~\ref{T:variables} outlines the four key variables specified for the proposed models. There are two outcome variables, the original label from Varol ($y_1$), as well as the \textit{cap.universal} score for each observation obtained from the Botometer API ($y_2$). These variables make a claim as to the level of automation of the observed account, which the models would attempt to predict using input vectors ($x$). The first input vectors ($x_1$) is the collection of network variables discussed in Table \ref{T:Network_Measures} in Section \ref{Sec:NetworkMeasures}. There is a total of 33 variables, consisting of 17 measures repeated for both the full network of each account, as well as the network result from k-core decomposition. The ego for any k-core decomposed network would always have an out-degree equal to k, which in this case is one. We therefore removed the out-degree centrality measurement from the decomposed network measures, resulting in a total of 33 measures. 

 \begin{table}[]
 \caption{Specified Variables}
 \label{T:variables}
 \begin{tabular}{@{}p{0.5cm}p{1.45cm}p{5.2cm}@{}}
 \toprule
 Code    & Variables     & Descripion                                            \\ \midrule
 $y_1$     & Binary        & Binary label from \citet{Varol2017}.                  \\
 $y_2$     & Continuous    & Cap.Universal score from Botometer.                   \\
 $x_1$     & Continuous    & 32 Network variables of full and reduced networks.    \\
 $x_2$     & Continuous    & Cap.Universal score from Botometer.                   \\ \bottomrule
 \end{tabular}
 \end{table}

\subsubsection{Ensemble Models}
Table \ref{T:modelformulas} outlines the four ensemble models. The first model (SL1) attempts to predict the outcome of the \citet{Varol2017} label by only using the proposed network measures as predictors. The second model differs only from the first by changeing trhe outcome variable from the Varol lables to the obtained scores from Botometer. The third model (SL3) isolates the Botometer score as a singular predictor vector with the \citet{Varol2017} label as the outcome. The final model (SL4) includes both the Botometer score and proposed network features as predictor with the original Varol labels.

The intention is to first compare models 1 and 3 to compare the isolated performance of the proposed features with Botometer. The second intention is to observe whether including the proposed feature with the scores from Botometer (SL4) would perform better than SL3. SL2 is simply for reference, to observe how well the proposed features would fare if predicting a continuous outcome as provided by Botometer.

\begin{table}[]
 \caption{Super Learner Ensemble Models}
 \label{T:modelformulas}
 \begin{tabular}{@{}p{0.72cm}p{1.45cm}p{5.2cm}@{}}
 \toprule
 Model & Formula          & Description\\ \midrule
 SL1   & $y_1\sim x_1$    &  Proposed feature on \citet{Varol2017} label. \\
 SL2   & $y_2\sim x_2$    &  Proposed feature on Botometer score. \\
 SL3   & $y_1\sim x_2$    &  Botometer on \citet{Varol2017} label. \\
 SL4   & $y_1\sim x_1+x_2$   &  Proposed feature combined with Botometer score on \citet{Varol2017} label. \\ \hline
 \end{tabular}
 \end{table}

With the four proposed ensemble models, and the data appropriately prepared, the base learner algorithms must be specified for the ensemble. The next section will briefly present the chosen algorithms.

\subsubsection{Specified Learners}
Each SL model was specified with a library of six learning algorithms, with an additional benchmarking algorithm. We include a benchmarking algorithm ``SL.mean'', which should be outperformed by more complex algorithms. The library includes a modest list of six algorithms: support vector machine \citep{Meyer2018} random forest \citep{Liaw2002}, cforest \citep{Hothorn2006,Strobl2007,Strobl2008}, ranger \citep{Wright2017}, and ibredbagg \citep{Peters2018}. The objective is to identify whether the proposed feature improves the performance of an existing approach to classification, and the interest is therefore in the difference in performance between the models and not the particular accuracy of a single model. It is, nevertheless, important to standardise the methodology of each model in order to ensure comparability.

Table \ref{T:algorithms} describes the estimated risk and individual weightings of the learner algorithms in each ensemble. The table also indicates where the hyper-parameters were altered from the defaults. The default parameters are those set in the SuperLearner package \citep{Polley2018}. For each model, an error distribution family must be specified, and a binomial family was specified for all models, except SL2, which has a continuous response variable, and is therefore specified a Gaussian family.

\begin{table*}[]
\caption{Ensemble Model Learner Libraries}
\label{T:algorithms}
\begin{tabular}{@{}lllll|llll|l@{}}
\toprule
\multirow{2}{*}{Algorithm} & \multicolumn{4}{c}{Estimated Risk} & \multicolumn{4}{c}{Coefficient} & \multirow{2}{*}{Model Hyper-parameters} \\ \cmidrule(lr){2-9}
 & SL1 & SL2 & SL3 & SL4 & SL1 & SL2 & SL3 & SL4 &  \\ \midrule
mean& 0.207 & 0.057 & 0.207 & \multicolumn{1}{l|}{0.207} & 0.000 & 0.000 & 0.000 & \multicolumn{1}{l|}{0.000} & NA \\
svm & 0.178 & 0.057 & 0.134 & \multicolumn{1}{l|}{0.136} & 0.056 & 0.000 & 0.166 & \multicolumn{1}{l|}{0.000} & defaults \\
cforest & 0.166 & 0.041 & 0.120 & \multicolumn{1}{l|}{0.113} & 0.455 & 0.044 & 0.454 & \multicolumn{1}{l|}{0.069} & defaults \\
xgboost & 0.202 & 0.045 & 0.121 & \multicolumn{1}{l|}{0.131} & 0.000 & 0.105 & 0.251 & \multicolumn{1}{l|}{0.068} & defaults \\
ranger & 0.168 & 0.040 & NA & \multicolumn{1}{l|}{0.135} & 0.000 & 0.249 & 0.000 & \multicolumn{1}{l|}{0.000} & num.trees=1000; mtry=2 \\
ipredbagg & 0.167 & 0.039 & 0.125 & \multicolumn{1}{l|}{0.111} & 0.283 & 0.529 & 0.128 & \multicolumn{1}{l|}{0.739} & nbagg=250 \\
randomForest & 0.168 & 0.040 & 0.140 & \multicolumn{1}{l|}{0.121} & 0.206 & 0.073 & 0.000 & \multicolumn{1}{l|}{0.124} & ntree = 3000 \\ \bottomrule

\end{tabular}
\end{table*}

Using an ensemble is not always guaranteed to offer a superior result over any single learner algorithm. It is therefore important to check the performance of each learner compared to the ensemble by comparing the estimated risk of each model.\footnote{See Section \ref{Sec:cv}}

\subsubsection{Model Performance Measures}
To measure performance of each model, we primarily use the area under the curve (AUC) metric derived from the return operator characteristic plot. We also highlight other measures, in particular to compare the performance of the models on different characteristics.

\section{Results}
To report the results of the applied methodology, we first have to investigate the validity of the specific ensemble approach by investigating the performance of the ensemble algorithm compared to the base algorithms. We then proceed to report the various appropriate accuracy measures of each proposed ensemble model.

\subsection{Ensemble Validity}
\label{Sec:cv}
\begin{figure}
    \centering
    \includegraphics[width=\linewidth]{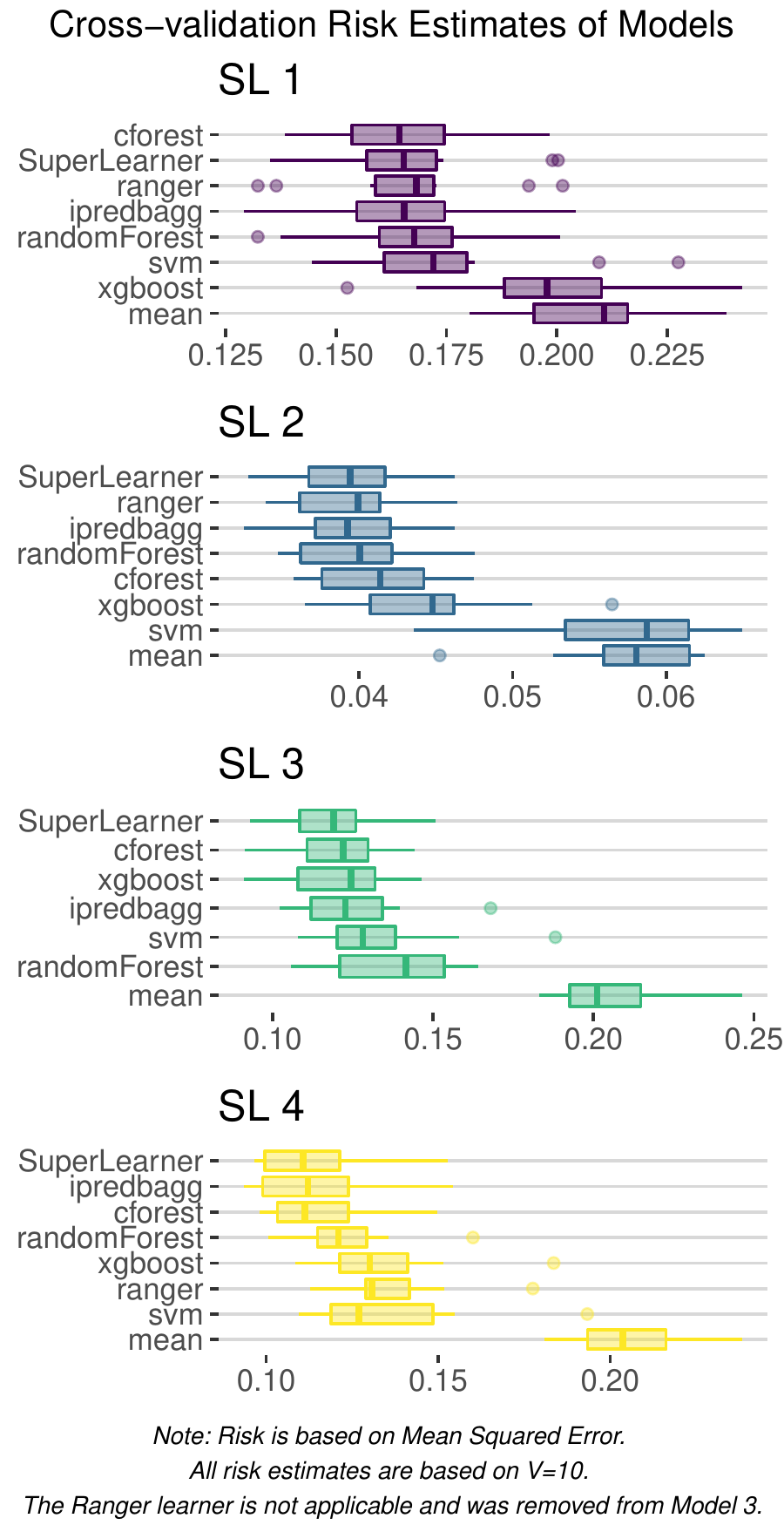}
    \caption{V-fold cross-validation results of SL library.}
    \label{fig:sl.cv}
\end{figure}

To determine whether the SL ensemble performs better than any single candidate algorithm, V-fold cross-validation was performed with ten folds. From Figure~\ref{fig:sl.cv} it is evident that the SL ensemble outperforms any single candidate algorithm, except in Model 1, where cforest has the lowest estimated risk. It is therefore beneficial to utilise ensemble learning with three of the four models. Each of the specified algorithms also outperformed the mean benchmark in each model. 

\subsection{Classification Performance}
To assess each model's performance, we highlight key classification performance measurements in Table \ref{T:performance}. We observe that SL4 narrowly outperforms SL3, and SL2 has the weakest performance in terms of AUC. This suggests that our intention to improve existing classification approaches by including our proposed features to traditional classification is a feasible objective. 

Another measure of interest is balanced accuracy. Balanced accuracy accounts for a disproportionate labelling in the training data \citep{Brodersen2010}. Our data does contain fewer cases of automated accounts than non-automated, which produce biased accuracy measures. Comparing the models on balanced accuracy reveals that SL4 offers a slight improvement over SL3. 

Considering precision and recall of the models, SL4 has the highest precision and SL3 has the highest recall. For an indication of balance between precision and recall, we use the F1 score, which indicates that SL3 is marginally higher than SL4.

\begin{figure}
    \centering
    \includegraphics[width=\linewidth]{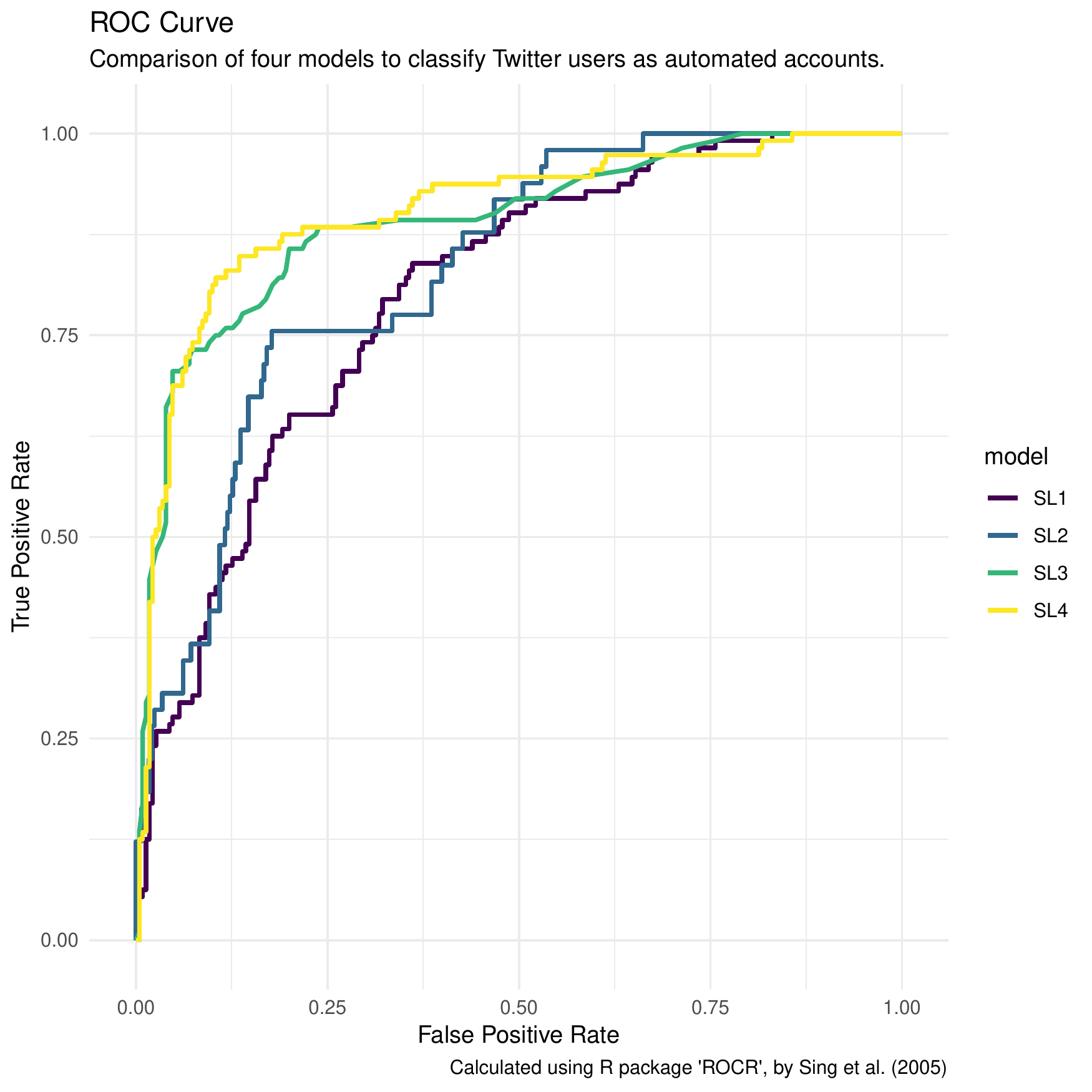}
    \caption{ROC Curve.}
    \label{fig:roc}
\end{figure}

\begin{figure}
    \centering
    \includegraphics[width=\linewidth]{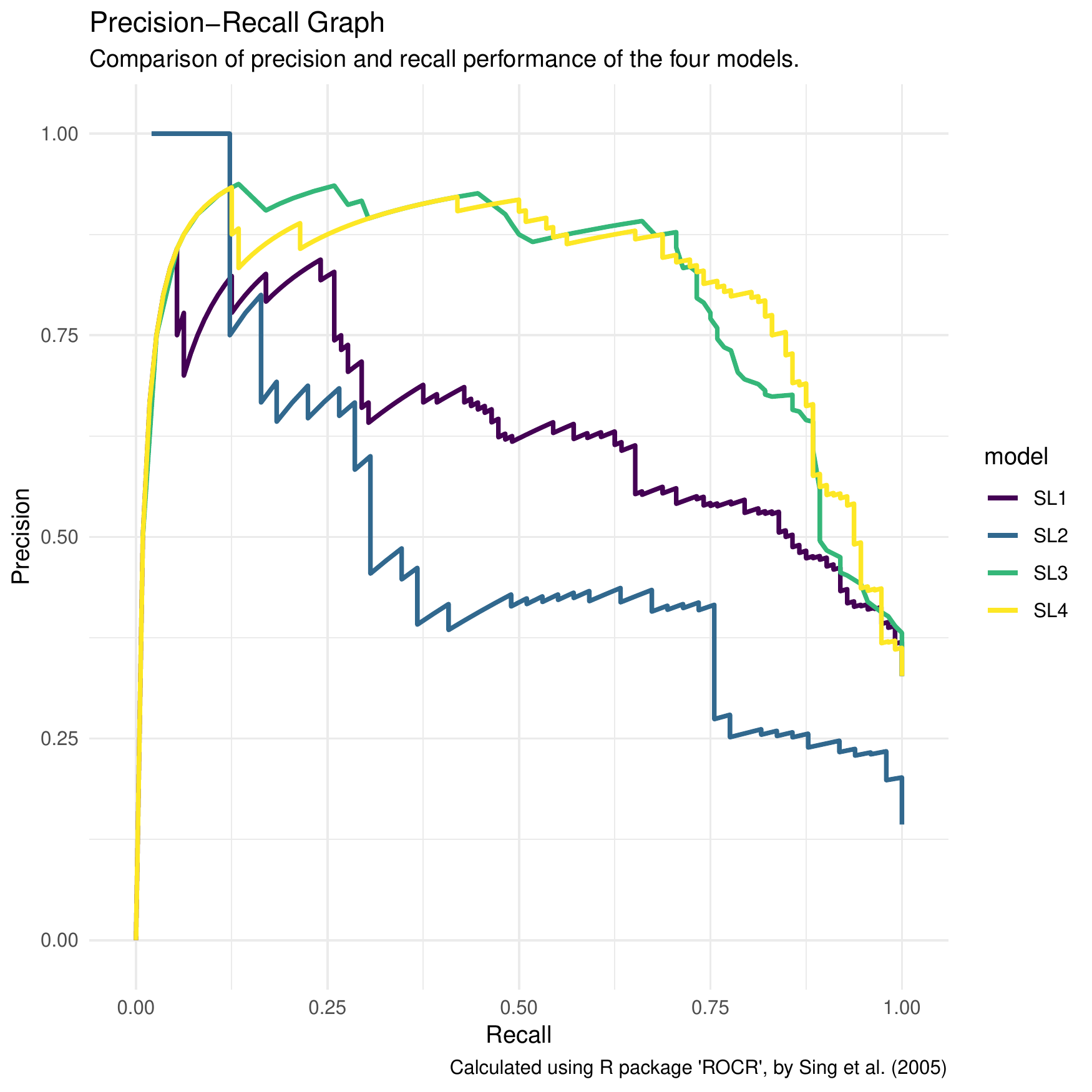}
    \caption{Precision-Recall Graph.}
    \label{fig:pr}
\end{figure}

All the performance measurements, except AUC, assume a single cutoff threshold. Plotting the results across a range of cutoff thresholds offers insights without making an assumption of a cutoff point. Figures \ref{fig:roc} and \ref{fig:pr} illustrate the performance of the models over all cutoff values. SL2 is predicted on a different outcome, and should therefore not be compared directly with the other models, particularly in Figure \ref{fig:pr}.

Figure \ref{fig:roc} makes it clear that SL3 and SL4 are dominant on both measures, with SL4 showing signs of improved robustness of false positives. Considering precision and recall in Figure \ref{fig:pr}, SL4 and SL3 are closely correlated, with SL4 providing higher precision with cutoffs above $\pm$0.7.

\begin{table}[]
\caption{Ensemble Model Performance Measures}
 \label{T:performance}
\begin{tabular}{lrrrr}
\toprule
Measure  & SL 1 & SL 2 & SL 3 & SL 4\\
\midrule
AUC & 0.799 & 0.830 & 0.891 & 0.903\\
Balanced Accuracy & 0.671 & 0.442 & 0.822 & 0.829\\
Precision & 0.771 & 0.250 & 0.870 & 0.879\\
Recall & 0.878 & 0.007 & 0.930 & 0.917\\
F1 & 0.821 & 0.013 & 0.899 & 0.898\\
\bottomrule
\end{tabular}
\end{table}

In conclusion, the results show an improvement of classification performance when combining a standard publicly available classifier (Botometer) and our proposed features. The next section will offer a discussion of these results, and elaborate on the key limitations and future work required.

\section{Discussion}
This paper proposes that current automated account classification techniques, regardless of success, can be improved upon by including a key missing feature of SNS accounts. Regardless of the results, there are limitations to this study in particular that needs explicit outlining. These limitations are usually opportunities for future studies to improve on, and we would therefore suggest future avenues of research.

\subsection{Socio-Informatic Approach to Bot Classification}
We propose that honest humans would be identifiable through their socio-informatic features online. Many \emph{dishonest} humans, and generally automated accounts, should exhibit distinct enough behaviour in contrast to the complex social signatures observed in fields such as social network analysis. 

To be sure, the creator of automated accounts would be able to set up a full network of accounts, which forms a convincing socio-informatic feature. This could be done by applying social heuristics such as a friend of a friend is a friend (increased clustering coefficient) or ensuring that the artificial network would have low average distance, while maintaining local high clustering---an archetype of observed social networks \citep[see][]{Kossinets2009}. Such fastidiousness by an individual to circumvent detection is almost amendable, but increasingly costly, and not fool proof. Such a creator can certainly control a bot-net to the triadic level, but any interface with honest networks would be extremely difficult and costly to establish, and should most certainly leave traces of unnatural social patterns.

Very few discerning social agents, with an honest use of SNS platforms such as Twitter would follow 6000+ accounts, unless, they have no information needs for following, or they use tools to filter the follower list to produce an actual curated information repertoire. Indeed, Twitter offers such a tool with `lists', but these are seldom used. It is tempting to propose follower cutoffs such as 150 or 250 followed accounts, in line with social brain hypothesis thinking \citep[see][]{Dunbar1998}. Instead of searching for such a cutoff, it might be more fruitful to combine informational and social observations of online accounts, which are filtered by complex social agents. Thus, the socio-informatic approach.

From the results of the study, we observe that the SL4 ensemble improves the classification performance of a leading classification service. This offers evidence that a more in-depth look into the more expensive socio-informatic behaviours as means of classifying social agents is a worthwhile venture. The particulars of where SL4 outperforms SL3 is also important. SL4 offers a more nuanced accuracy, particularly more precision on classification. SL4 also outperforms SL3 at higher thresholds, which suggests that it makes less mistakes when `pushed' for recall of automated accounts. It might therefore prove to offer a more robust feature set, resilient to \textit{faux} socio-informatic behaviour of online accounts.

Regardless of the suggested success of this approach, there are some key limitations that warrants a reserved conclusion. The next section will elaborate on the identified limitations, before concluding the paper.

\subsection{Limitations}

A machine learning model's real-world prediction performance is wholly reliant on the quality of the ground truth on which it is trained. In this paper, we relied on the manual annotations provided by \citet{Varol2017}. Due to the nature of the classification task, even manual annotation by experts can prove to be inaccurate since there is no definitive manner by which an annotation can be confirmed \citep[see][]{Wang2012}. We can highlight examples of potential miss-classifications. After obtaining the two-step follower network, and calculating the network measurements, we observe duplicate measures in the data. Upon inspection, the duplication of the network measures turn out to be exactly the same network structure between two different accounts. This could be due to very small networks containing only two nodes, but this was not the case. From the full networks, an example is two accounts with identical networks (apart from the ego node) which have a size of 53067 nodes, where the one is labelled as automated and the other non-automated. From the reduced networks, these duplicates become more prevalent, and most are due to small size. Investigating those with size 0 reveals that most are labelled as automated, but some are labelled as not. Investigating these non-automated accounts reveals that these accounts are trivial CRM accounts, who follow a single user. The social richness of these accounts are therefore poor, but not `dishonest'. 

The second limitation to this approach is that the data collection process is slow. To collect the follower list of a single user is fast, but iterating over that list to return each followed user's list of followed accounts increases the collection time dramatically. The Twitter API is the key bottleneck in the process, since it restricts the amount of calls allowed within a certain time-frame. This makes this approach impractical for fast classification of unseen accounts, such as required by a service such as Botometer. However, the process should prove to be less restrictive for the owners of the platform itself. It is also feasible to maintain an independent database of observed networks with periodical updates, in order to provide faster classification. 

The third limitation is again due to the nature of the \citet{Varol2017} dataset, however, it is a limitation of any such publicly available dataset. Twitter actively suspends accounts which break its terms and conditions, and accounts such as these are captured in the datasets. The attrition rate for automated accounts is therefore high and unpredictable. We therefore have an unbalanced dataset, containing more examples of non-automated accounts than automated. Although we attempt to account for this unbalanced representation, it is not ideal.

The final limitation is the na\"{i}ve approach of only combining Botometer scores and our proposed features. It would be preferable to include all 1000 features as used by Botometer and the 33 features introduced here in a machine learning model. However, the exact features and machine learning parameters of Botometer are not known. We therefore opted for an intuitive, yet na\"{i}ve approach of simply combining the predictions from Botometer as a proxy for their full feature set.

\subsection{Future Research}
Future research would be able to offer further insights of the viability and value of a socio-informatic feature of in SNS account classification. The current literature on classification within this domain is already moving away from binary classification assumptions. For instance, there are accounts with varying degrees of automation, social complexity and honesty. It is perfectly reasonable to expect automated accounts to be honest (client relations management accounts), and non-automated accounts to be dishonest (sock-puppets). It is therefore important to extend the socio-informatic focus to aid in untangling rich social agents from poor social agents.

The first steps in such an approach is to increase and broaden the underlying data to include better labels for training, or potentially continuum's of the quality of social agents. Such nuanced classifications could potentially benefit more from a socio-informatic approach, however it could prove cumbersome in the case of binary classification. 

Further research should also attempt a model-level integration of socio-informatic and classic features. The inclusion of profile information, temporal features and sentiment of profiles within the model might return better results than using the na\"{i}ve approach employed here. It is also possible, yet more challenging, to collect the same information repertoire networks at different intervals to observe changes in the network as further evidence of active social pruning and adaptation of the network of the user.

\section{Conclusion}
The study proposed that capturing a socio-informatic feature of a user on a SNS would provide a valuable addition in machine learning classification attempts of automated agents. The socio-informatic feature was operationalised as a two step follower network on a popular SNS platform (Twitter). These networks were analysed using various social network measurements, which was then included in an ensemble learning approach to automated agent classification. We also included a third party classification on the same dataset. The classification results of the third-party classifier was compared with a model which added the proposed features. It was found that the addition of the proposed socio-informatic feature improved the predictive performance of the model. 
%
\bibliographystyle{ACM-Reference-Format}
\bibliography{references2}

%

\end{document}